# Bias-Independent Subthreshold Swing in Nanoscale Cold-Source Field-Effect Transistors by Drain Density-of-States Engineering

Kunyi Liu, Fei Lu, and Yuan Li*

*School of Information Science and Engineering, Shandong University, Qingdao 266237, China.*
*E-mail: yuan.li@sdu.edu.cn

We report a strategy to design nanoscale cold-source field-effect transistors (CS-FETs) with bias-independent sub-60 mV/dec subthreshold swing (SS). By first-principles calculations and quantum-transport simulations, we reveal that the energy alignment of density of states (DOS) between the drain and source electrodes is critical to achieving bias-independent SS. By defining "gate window", we propose a device model to demonstrate how similar slopes of the drain DOS falling into the gate window can stabilize the SS under different bias. This study underscores the significance of drain DOS engineering in the design of CS-FETs with bias-independent SS for portable electronic applications.

With scaling down field-effect transistors (FETs) toward the physical limit, power dissipation of highly-integrated FETs has become a serious issue impeding the advancement of integrate circuits.[1] The wide variation in circuit characteristics caused by power supply also impacts the performance of low-power digital circuits.[2] Since the power consumption and the performance of FETs/circuits are both sensitive to the power supply, it is useful to design in a way such that the supply voltage remains as stable as possible. However, such requirement fails to be fulfilled in many applications, such as the edge-computing devices/systems driven by low-power portable battery, of which the supply voltage degrades over time due to the limited power budget.[3] As such, an alternative strategy is to optimize the relevant device parameters, such as subthreshold swing (SS), to be independent of the supply voltage as much as possible. It is noted that SS, defined as the variation of gate voltage per change of leakage current by an order of magnitude, represents a key parameter governing both power consumption and performance of FETs.[4] Small values of SS are required in order to achieve low-power and high-performance FETs in the subthreshold regime. However, the SS of FETs usually suffers from an inherent lower limit (about 60 mV/dec at 300 K) known as the "Boltzmann tyranny", which is caused by the electronic thermionic tail of the device electrodes. To break this limit, cold-source field-effect transistors (CS-FETs) were recently proposed by employing the so-called "cold" electron source, of which the thermionic tail is suppressed by the exponential decay of the electron density to achieve sub-60 mV/dec SS.[5] CS-FETs were intensively investigated in recent studies by proposing more "cold" materials and/or optimized architectures to obtain smaller SS.[4,6-8] However, most CS-FETs display explicit dependence of the SS on the bias voltage and the underlying mechanism, which is essential for optimizing the CS-FETs in practical applications, still remains unclear.[7]

In this work, by first-principles calculations and quantum-transport simulations, we propose a strategy to design nanoscale CS-FETs with bias-independent sub-60 mV/dec SS by engineering the density of states (DOS) of drain electrode, which was rarely focused in previous studies of CS-FETs. We reveal that, for a given bias voltage, the energy alignment of DOS between the drain and source electrodes plays a dominant role in impacting the SS of CS-FETs. In terms of CS-FETs with given source and channel materials, drain DOS engineering, as will be shown, plays a critical role in decoupling the dependence of SS on the bias voltage.

We start with an analysis on how the SS of CS-FETs can be impacted by bias voltage. Taking p-type CS-FETs as an example, we proposed a model of the devices with different characteristics of drain DOS, as shown schematically in Fig. 1. In this model, we focused on the source and drain DOS around the electrode Fermi levels as well as their energy alignment bridged with the valence band edge (VBE) of the channel semiconductors. The DOS with linear and regular energy dependence were employed to model the cold source and the regular drain electrodes, respectively. By taking the Fermi level $E_{\text{fs}}$ of the source electrode as energy reference, the VBE can be shifted up or down with respect to $E_{\text{fs}}$ when a gate voltage is applied, while the Fermi level $E_{\text{fd}}$ of the drain electrode is lifted when applying a forward bias voltage. For nanoscale CS-FETs of interest in this work, ballistic quantum transport dominates such that the electric current of the devices can be determined by the Landauer-Büttiker formula:[9]

$$I = \frac{2e}{h}\int_{-\infty}^{+\infty}\{T(E)[f_s(E-E_{\text{fs}}) - f_d(E-E_{\text{fd}})]\}dE \quad (1)$$

where $T(E)$ denotes the transmission coefficient as a function of energy $E$; $f_{s/d}(E-E_{\text{fs/d}})$, the Fermi-Dirac distribution function at equilibrium of the source/drain electrode; $e$, the elementary charge; and $h$, the Planck's constant. According to Eq. (1), the energy-resolved electric current (spectral current) is proportional to the transmission coefficient, while the latter is also proportional to the product of the DOS projected onto the source ($g_s(E)$) and drain ($g_d(E)$) electrodes,[10] i.e.,

$$I(E) \propto T(E) \propto g_s(E)g_d(E) \quad (2)$$

For a given bias voltage, $g_{s/d}(E)$ remains unchanged in energy such that the electric current is governed by the alignment between $g_s(E)$ and $g_d(E)$ bridged with the channel. When modulating the gate voltage within a certain range, such as from $V_{\text{gs1}}$ to $V_{\text{gs2}}$ as shown in Fig. 1, it is useful to define a "gate window" to measure the energy difference between the VBE of the channel semiconductors under the different gate voltages. In this context, the SS calculated within the gate window is essentially determined by the difference between (rather than the magnitude of) the corresponding electric currents, which, according to Eq. (2), are proportional to $g_d(E)$ for a given source electrode. As such, it is expected to achieve different SS when different slopes of the drain DOS fall into the gate window. For instance, larger SS can be expected in the case of small bias with downward slope of the drain DOS within the gate window (see Fig. 1(a)), when compared with the case of large bias with upward slope of the drain DOS within the gate window (see Fig. 1(b)). Otherwise, however, comparable SS can be expected when the slopes of the drain DOS within the gate window are similar (see Figs. 1(c) and 1(d)). Because of its generality, this principle is also expected to work in other nanoscale FETs in which ballistic quantum transport dominates.

Following the analysis discussed above, we designed double-gate p-type nanoscale CS-FETs, of which the geometry has been displayed in Fig. 2(a). Here, we employed monolayer $C_{31}$, a two-dimensional (2D) carbon allotrope that was recently applied in CS-FET,[11] as the cold-source electrode and 2D $MoS_2$ as the channel and the drain electrode (via doping), respectively. The geometric structure of the $C_{31}$ lattice has been presented in Fig. 2(b). First-principles calculations at the density functional theory (DFT) level were employed to optimize the geometry of the materials and devices as well as to estimate the electronic properties. In practice, we employed the exchange-correlation functional at the generalized-gradient approximation Perdew-Burke-Ernzerhof (GGA-PBE) level and the PseudoDojo potential with energy cutoff of 680 eV. The Brillouin zone was sampled by $12\times12\times1$ Monkhorst-Pack $k$-points, and the force tolerance was set as 0.001 eV/Å. A vacuum space of 2 nm was applied along the $z$ direction to avoid the interaction error of periodic images. The quantum-transport properties of the CS-FETs were calculated in frame work of DFT combined with nonequilibrium Green's function method.[12] The electric current of the devices was calculated according to the Landauer-Büttiker formula (Eq. (1)). In practice, the $k$-point grids of $3\times150\times1$ and $12\times12\times1$ were employed for the self-consistent calculations and the transmission analysis, respectively. The periodic, Neumann, and Dirichlet boundary conditions were applied along the transverse, vertical, and transport directions, respectively. The mesh cutoff for real space was set as 105 Hartree. The temperature used for all the device simulations was 300 K. All the calculations were performed by using the Quantum ATK package.[13,14]

Before evaluating the device performance, we first calculated the electronic band structures and DOS of a $C_{31}$ monolayer as well as a van der Waals (vdW) heterojunction of $C_{31}/MoS_2$. The corresponding results have been illustrated in Figs. 2(c) and 2(d), respectively. As shown in Fig. 2(c), the $C_{31}$ monolayer displays semi-metallic band structures and DOS, which is consistent with the results of previous studies.[11] The electronic states around the Fermi level have major contributions from the $p_z$ orbitals of carbon atoms. The DOS above the Fermi level manifests itself with a set of isolated states, which were demonstrated to be able to self-filter out thermionic tails such that $C_{31}$ monolayer can serve as cold-source electrode for p-type CS-FETs.[11] When contacting with 2D $MoS_2$ via vdW interaction, as shown in Fig. 2(d), the $C_{31}$ monolayer maintains its electronic properties in spite of a slightly downward shift of the Fermi level because of electron transfer at the heterojunction interface. In the CS-FETs studied here (see Fig. 2(a)), the 2D $MoS_2$ remains intrinsic when it serves as the channel, while it is p-type doped to make use of its peculiar DOS around the top of valence band when it serves as the drain electrode (see Fig. 2).

To evaluate the device performance, we calculated the transfer characteristics of the CS-FETs, i.e., the electric current $I$ as a function of the gate voltage $V_g$ under different bias $V_{ds}$. The results in the cases of heavily-doped (with doping concentration of $5.4 \times 10^{20}$ cm$^{-3}$) and lightly-doped (with doping concentration of $1.4 \times 10^{20}$ cm$^{-3}$) drain electrodes have been illustrated in Figs. 3(a) and 3(b), respectively. We need to point out that, to better display the results, each $I - V_g$ curve was transversely shifted by the corresponding bias voltage $V_{ds}$, i.e., $V_g = V_{gs} + V_{ds}$. It is seen that all the curves present transfer characteristics typical of the subthreshold regime, i.e., $I$ increases exponentially as a function of increasing $V_g$. The SS was calculated by using the steepest part of each $I - V_g$ curve (see the gray arrows in Figs. 3(a) and 3(b)). The results of SS as a function of the bias voltage in the cases of both light and heavy doping have been displayed in Fig. 3(c). It is seen that SS with sub-60 mV/dec was obtained for all the $I - V_g$ curves, indicating that the thermionic tails are indeed suppressed by the C$_{31}$ cold source in the devices. It is also important to note that the SS decreases (from about 53 to 23 mV/dec) with increasing the bias voltage in the case of heavy doping, while it remains almost constant (around 37 mV/dec) in the case of light doping. To verify these results, we have potted in Fig. 3(d) an enlarged view of the steepest part of the $I - V_g$ curves for both cases of the heavy and light doping under small bias of 0.2 V and large bias of 0.6 V, respectively. It is obvious that the $I - V_g$ curve under $V_{ds} = 0.6$ V is steeper than that under $V_{ds} = 0.2$ V in the case of heavy doping, while they are almost as steep as one another in the case of light doping. Therefore, bias-independent SS is indeed obtained in the designed CS-FETs with lightly-doped drain electrode.

To reveal the underlying mechanism, we calculated the projected DOS of the source, channel, and drain corresponding to the data shown in Fig. 3(d), and the results have been presented in Fig. 4. Following the analysis based on Fig. 1, we plotted the gate windows (the regions between the dotted lines shown in Fig. 4) as a result of the gate voltage change from $V_{gs} = -3.2$ V to $-3.3$ V. The bias windows (the regions between the dashed lines $E_{fs}$ and $E_{fd}$ shown in Fig. 4) were also plotted to visualize the bias voltage change from $V_{ds} = 0.2$ V to 0.6 V. Since the gate windows are out of the bias windows, the gate voltages mostly modulate the thermal rather than tunneling component of the electric currents, which is consistent with the operation of CS-FETs. As such, under given bias voltage, the SS of the devices is essentially determined by the energy alignment between the DOS of the source and drain electrodes bridged with the channel. To be more specific, the SS is governed by the slope of the drain DOS falling into the gate window because that of the source and channel remains bias independent. For instance, in the case of heavy doping (see Figs. 4(a) and 4(b)), the slope of the drain DOS under small bias is downward with increasing energy (see Fig. 4(a)), while it is just the opposite under large bias (see Fig. 4(b)). As a result, the SS obtained under large bias is smaller than that under small bias because of the different slopes of the drain DOS. This is consistent with the analysis based on Figs. 1(a) and 1(b). On the other hand, in the case of light doping (see Figs. 4(c) and 4(d)), the slopes of the drain DOS under both small and large bias are very similar, i.e., both downward with increasing energy. As a result, the SS remains almost independent of the bias voltage, which also agrees well with the analysis based on Figs. 1(c) and 1(d). These results strongly suggest that drain DOS engineering to obtain similar slopes of DOS falling into the gate window is of critical importance for decoupling the dependence of SS with bias voltage in the CS-FETs. As a matter of fact, it is technically available in experiments to engineer the DOS by means of doping. For instance, chloride molecules were demonstrated experimentally to be effective in doping 2D $MoS_2$;[15] cesium carbonate surface functionalization was also successfully applied to achieve stable doping of monolayer $MoS_2$.[16] Therefore, it is experimentally possible to realize bias-independent SS of CS-FETs by appropriately engineering the DOS of drain electrode.

In conclusion, we have reported a strategy to design CS-FETs with sub-60 mV/dec SS independent of the bias voltage, which is required for portable electronic applications. By first-principles calculations and quantum-transport simulations, we designed and characterized nanoscale CS-FETs based on a vdW heterojunction of 2D $C_{31}/MoS_2$ and obtained bias-independent sub-60 mV/dec SS when the $MoS_2$ drain electrode is lightly doped. By proposing a device model, we reveal that the bias-independent SS is essentially determined by the energy alignment between the drain and source DOS. In particular, engineering of the drain DOS with similar slopes falling into the gate window was demonstrated to be important for achieving bias-independent SS. This study provides an insight into how the SS is impacted by the bias voltage in nanoscale CS-FETs as well as a guideline for the design of more robust low-power FETs.

**Acknowledgments**

This work was supported by the National Natural Science Foundation of China (Nos. 62174100 and 61874068).

# References

1) A. M. Ionescu and H. Riel, Nature **479**, 329 (2011).
2) N. Verma, J. Kwong and A. P. Chandrakasan, IEEE Trans. Electron. Dev. **55**, 163 (2008).
3) A. Tajalli and Y. Leblebici, IEEE Trans. Circuits Syst. II, Exp. Briefs **56**, 374 (2009).
4) F. Liu, C. Qiu, Z. Zhang, L. Peng, J. Wang, and H. Guo, IEEE Trans. Electron Devices **65**, 2736 (2018).
5) C. Qiu, F Liu, L. Xu, B. Deng, M. Xiao, J. Si, L. Lin, Z. Zhang, J. Wang, H. Guo, H. Peng and L. Peng, Science **361**, 387 (2018).
6) J. Lyu, J. Pei, Y. Guo, J. Gong and H. Li, Adv. Mater. **32**, 1906000 (2020).
7) Q. Wang, P. Sang, X. Ma, F. Wang, W. Wei, W. Zhang, Y. Li and J. Chen, IEEE Int. Electron Devices Meet, 2020, p. 22.4.1.
8) P. Sang, Q. Wang, W. Wei, L. Tai, X. Zhan, Y. Li and J. Chen, IEEE Trans. Electron Devices **69**, 2173 (2022).
9) M. Büttiker, Y. Imry, R. Landauer, and S. Pinhas, Phys. Rev. B **31**, 6207 (1985).
10) Y. Pan, Y. Wang, L. Wang, H. Zhong, R. Quhe, Z. Ni, M. Ye, W. Mei, J. Shi, W. Guo, J. Yang and J. Lu, Nanoscale **7**, 2116 (2015).
11) Q. Wang, P. Sang, F. Wang, W. Wei, Y. Li and J. Chen, Appl. Phys. Express **14**, 074003 (2021).
12) M. Brandbyge, J. Mozos, P. Ordejón, J. Taylor, and K. Stokbro, Phys. Rev. B **65**, 165401 (2002).
13) QuantumATK version Q-2019.12, Synopsys QuantumATK (Available: https://www.synopsys.com/silicon/quantumatk.html).
14) S. Smidstrup et al., J. Phys.: Condens. Matter. **32**, 015901 (2020).
15) L. Yang, K. Majumdar, H. Liu, Y. Du, H. Wu, M. Hatzistergos, P. Y. Hung, R. Tieckelmann, W. Tsai, C. Hobbs, and P. D. Ye, Nano Lett. **14**, 6275 (2014).
16) J. D. Lin, C. Han, F. Wang, R. Wang, D. Xiang, S. Qin, X. Zhang, L. Wang, H. Zhang, A. T. S. Wee, and W. Chen, ACS Nano **8**, 5323 (2014).

**Figure Captions**

**Fig. 1.** (Color online) Schematic diagram of a device model of p-type CS-FETs with different characteristics of drain density of states (DOS). (a) and (b) The drain DOS within the gate window evolves non-monotonously as a function of energy when changing from (a) small to (b) large bias voltages. (c) and (d) The drain DOS within the gate window evolves monotonously as a function of energy when changing from (c) small to (d) large bias voltages. The gate window refers to the energy difference between the valence band edge of the channel semiconductors under the different gate voltages ($V_{gs1}$ and $V_{gs2}$). $E_{fs(d)}$ denotes the Fermi level of the source (drain) electrode. The arrows indicate the slope of the DOS within the gate windows.

**Fig. 2.** (Color online) (a) Geometric structure of a p-type CS-FET with 2D $C_{31}$ as source and $MoS_2$ as channel and drain. The gate length is $L = 5.9$ nm. (b) Top and side views of the optimized geometry of a $C_{31}$ monolayer. (c) Atomic-orbital-resolved electronic band structures (left panel) and DOS (righ panel) of a $C_{31}$ monolayer. (d) Element-resolved electronic band structures (left panel) and DOS (right panel) of a $C_{31}/MoS_2$ van der Waals heterojunction. The Fermi level is at zero energy.

**Fig. 3.** (Color online) Transfer characteristics of the $C_{31}/MoS_2$-based CS-FETs with the drain part of the $MoS_2$ p-type (a) heavily doped (with concentration of $5.4 \times 10^{20}$ cm$^{-3}$) and (b) lightly doped (with concentration of $1.4 \times 10^{20}$ cm$^{-3}$), respectively. The gray arrows denote the steepest part of the curves where the subthreshold swing (SS) was calculated. (c) SS as a function of the bias voltage obtained for the heavily-doped and lightly-doped devices, respectively. (d) Enlarged view of the steepest part of the transfer characteristics for the heavily-doped and lightly-doped devices under bias voltages of 0.2 V and 0.6 V, respectively.

**Fig. 4.** (Color online) (a) and (b) Projected DOS of the $C_{31}/MoS_2$-based CS-FET when the drain region of $MoS_2$ is p-type heavily doped (with concentration of $5.4 \times 10^{20}$ cm$^{-3}$) under (a) small bias voltage (0.2 V) and (b) large bias voltage (0.6 V). (c) and (d) Projected DOS of the $C_{31}/MoS_2$-based CS-FET when the drain region of $MoS_2$ is p-type lightly doped (with concentration of $1.4 \times 10^{20}$ cm$^{-3}$) under (c) small bias voltage (0.2 V) and (d) large bias voltage (0.6 V). The source Fermi level is set as $E_{fs} = 0$. The regions between the dotted lines denote the gate windows, while those between the dashed lines ($E_{fs}$ and $E_{fd}$) represents the bias windows. The arrows denote a guide-to-the-eye of the DOS slopes within the gate windows.

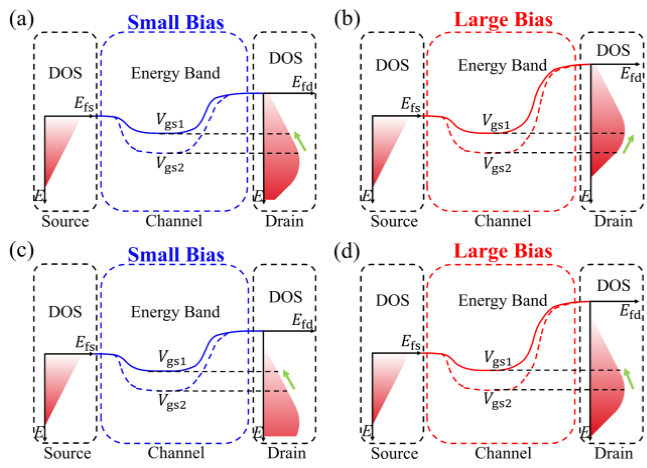

Fig.1.

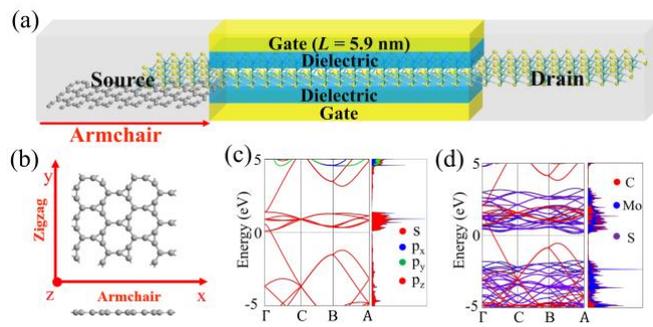

Fig. 2.

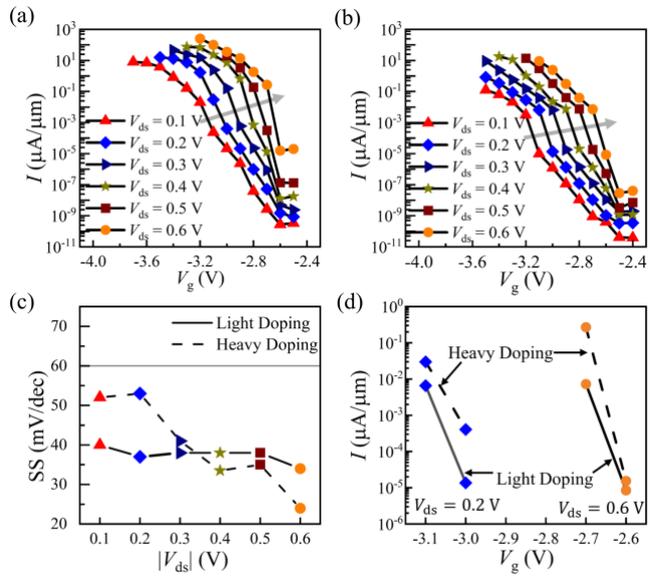

Fig. 3.

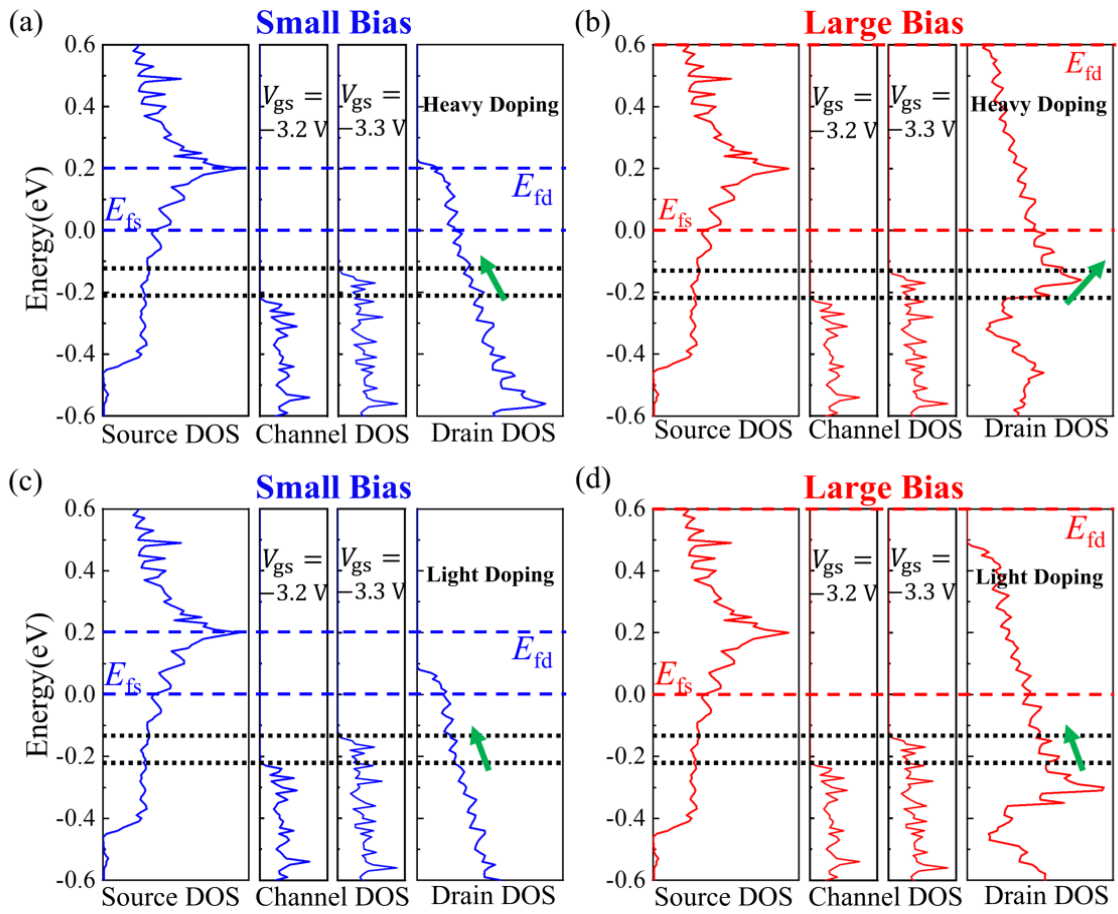

Fig. 4.